\documentclass[letterpaper, 10 pt, conference]{ieeeconf}  

\IEEEoverridecommandlockouts
\usepackage{amsmath,amssymb,amsfonts}
\usepackage{graphicx}
\usepackage{textcomp}
\usepackage{graphics} 
\usepackage{float}
\usepackage{siunitx}
\usepackage{cite}

\usepackage{jabbrv}

\usepackage[caption=false, font=footnotesize]{subfig}

\usepackage{tikz}
\usetikzlibrary{arrows.meta,shadows,positioning,shapes.geometric,intersections}

\tikzset{
	frame/.style={
		rectangle, draw,
		text width=6em, text centered,
		minimum height=3em,drop shadow,fill=white,
		},
		line/.style={
			draw, -{Latex},
          },
 }

\title{\LARGE \bf{Safe reinforcement learning control for continuous-time nonlinear systems without
    a backup controller}}
\author{Soutrik Bandyopadhyay and Shubhendu Bhasin
\thanks{The authors are with the Department of Electrical Engineering, Indian Institute of Technology Delhi, New Delhi, India.
        {\tt\small \{Soutrik.Bandyopadhyay,sbhasin\}@ee.iitd.ac.in}}
}

\newtheorem{thm}{Theorem}
\newtheorem{examp}{Example}
\newtheorem{construction}{Construction}

\newtheorem{rem}{Remark}
\newtheorem{assm}{Assumption}
\newtheorem{prob}{Problem}

\newcommand{\C}{\mathcal{C}}

\newcommand{\R}[2]{\mathbb{R}^{#1 \times #2}}
\renewcommand{\Re}{\mathbb{R}}
\renewcommand{\quad}[2]{#2^T #1 #2}

\newcommand{\Wf}{W_f}
\newcommand{\WfT}{W_f^T}
\newcommand{\Vf}{V_f}
\newcommand{\VfT}{V_f^T}

\newcommand{\WfHat} {\hat{W}_f}
\newcommand{\WfHatT}{\hat{W}_f^T}
\newcommand{\VfHat} {\hat{V}_f}
\newcommand{\VfHatT}{\hat{V}_f^T}

\newcommand{\sig}{\sigma(\VfT x)}
\newcommand{\sigHat}{\sigma(\VfHatT x)}

\newcommand{\ef}{\epsilon_f}
\newcommand{\ev}{\epsilon_v}

\newcommand{\Linf}{\mathcal{L}_{\infty}}

\newcommand{\WfTilde}{\tilde{W}_f}
\newcommand{\VfTilde}{\tilde{V}_f}

\newcommand{\Wfquad}{\WfTilde^T \Gamma_{wf}^{-1} \WfTilde}
\newcommand{\Vfquad}{\VfTilde^T \Gamma_{vf}^{-1} \VfTilde}

\newcommand{\gradPhi}{\nabla \phi}
\newcommand{\gradPhiT}{\nabla \phi^T}
\newcommand{\gradBf}{\nabla B_f}

\newcommand{\norm}[1]{\|#1\|}

\newcommand{\WcHat}{\hat{W}_c}
\newcommand{\WaHat}{\hat{W}_a}

\newcommand{\WaTilde}{\tilde{W}_a}
\newcommand{\WcTilde}{\tilde{W}_c}

\newcommand{\WcTildeDot}{\dot{\tilde{W}}_c}

\newcommand{\delhjb}{\delta_{hjb}}

\newcommand{\WaBar}{k_a}
\newcommand{\RsBar}{k_s}
\newcommand{\ToneBar}{k_1}
\newcommand{\TtwoBar}{k_2}
\newcommand{\fall}{\forall}

\newcommand{\ifig}[2]{\includegraphics[width=#1\linewidth]{img/FinalPlots/#2.png}}

\DeclareMathOperator*{\argmin}{arg\,min}

\DeclareMathOperator{\vect}{vec}
\DeclareMathOperator{\proj}{proj}

\usepackage{subfiles}

\makeatletter
\let\NAT@parse\undefined
\makeatother
\usepackage{hyperref}  
\hypersetup{
    colorlinks=true,
    linkcolor=blue,
    filecolor=magenta,
    urlcolor=cyan,
    citecolor=red,
    }
\begin{document}

\setlength{\abovedisplayskip}{5pt}
\setlength{\belowdisplayskip}{5pt}

\maketitle
\thispagestyle{empty}
\pagestyle{empty}

\begin{abstract}
This paper proposes an on-policy reinforcement learning (RL) control algorithm
that solves the optimal regulation problem for a class of uncertain
continuous-time nonlinear systems under user-defined state constraints. We
formulate the safe RL problem as the minimization of the Hamiltonian subject to
a constraint on the time-derivative of a barrier Lyapunov function (BLF). We subsequently use the analytical solution of the optimization problem to modify the Actor-Critic-Identifier architecture to learn the optimal control policy safely. The proposed method does not require the presence of external backup controllers, and the RL policy ensures safety for the entire duration. The efficacy of the proposed controller is demonstrated on a class of Euler-Lagrange systems.
\end{abstract}

\section{Introduction}
\label{sec:intro}

The reinforcement learning (RL) framework has seen reasonable success in solving
optimal control problems under uncertain system dynamics. However, most RL-based
methods need to explore the state-action spaces during the initial phases of
training. Consequently, they tend to apply control inputs that may
be detrimental to real-time safety-critical systems. This fundamental challenge
of RL algorithms precludes their use in real-world systems lest they endanger
the safety of humans and property. Therefore, researchers actively seek to
bolster RL algorithms with provable safety guarantees. Formally, the notion of
safety of dynamical systems is the certification of forward
invariance\cite{blanchini1999Automatica} of state and actuation constraint sets.
Under this definition of safety, the safe RL problem is the
mathematical construct to solve optimal control problems under user-defined
state and actuation constraints.

In literature, various methods are proposed to ensure the safety of RL
algorithms. One school of
thought is to exploit model predictive control (MPC) to buttress
RL algorithms with safety guarantees\cite{zanon2020TAC,wabersich2021TAC,li2018ACC}. While
these algorithms provide a unified approach to handling state and actuation
constraints, they solve an optimization routine at each
time step of the controller run and thus are computationally expensive.

Another class of methods in the safe RL literature employs control
barrier functions (CBF)\cite{ames2016TAC,ames2019ECC}. CBFs provide a
Lyapunov-like analysis to ensure the safety of dynamical systems
without the need to compute system trajectories. In
literature, it is common to combine CBFs with control Lyapunov functions (CLFs) in the form of an
optimization problem to
trade-off safety and stability objectives\cite{choi2020Arxiv}.
However, these approaches are limited to discrete-time control
problems.

The extension of the results of RL to uncertain continuous-time systems have
been achieved by combining approximate dynamic programming
(ADP) with adaptive control
\cite{vamvoudakis2017SysConLet,bhasin2013Automatica,vamvoudakis2010Automatica,lewis2009CircuitsSystemsMagazine}.
These approaches approximately solve the unconstrained optimal control
problem for uncertain system dynamics, however, the constrained optimal control
problems for continuous-time systems remain an active area of research.

The safety problem of continuous-time RL is primarily addressed by
considering the continuous-time counterpart of CBFs, namely barrier Lyapunov
functions (BLF) \cite{tee2009Automatica}. One research direction is to
transform the constrained state dynamics into dynamics of an unconstrained
state \cite{yang2019ACC,greene2020LCSS,mahmud2021ACC} and subsequently use ADP
algorithms to solve the unconstrained problem. However, this approach
typically handles rectangular state constraints (box constraints on individual
components of states) and cannot be trivially extended to general convex state constraints.
Additionally, these approaches modify the original cost function non-trivially.

Another approach involves adding BLF to the cost formulation
\cite{marvi2021IJRNC,cohen2020CDC}. Such an addition often renders the
system's value function not continuously differentiable, which is typically
needed to establish theoretical guarantees of the algorithms.

A common feature in both continuous-time and discrete-time RL algorithms is
the use of the so-called ``backup controllers''\cite[Assm.
2]{mahmud2021ACC}\cite{almubarak2021CDC}. These are user-defined
stabilizing controllers that step in place when RL algorithms generate control
actions not in accordance with the safety requirements. Most literature assumes
access to an initial policy that stabilizes the system under a wide range of
epistemic uncertainties. The backup controllers are typically used as a
fallback measure during the initial phase of the RL training when the agent has
limited knowledge of the system under control. The assumption of the
availability of such controllers is restrictive, and the formulation of
backup controllers may be difficult for certain complex systems. Additionally,
the act of switching to a backup controller deviates from the on-policy RL algorithm leading to sub-optimal results.

In this paper, an on-policy RL algorithm is developed for the optimal control of
continuous-time nonlinear systems that guarantee safety while obviating the need
for a backup controller.
Furthermore, the
objective function of the optimal control problem remains unchanged. Inspired by\cite{almubarak2021CDC}, we focus our efforts on extending the
Actor-Critic-Identifier (ACI) architecture\cite{bhasin2013Automatica} to solve
the optimal regulation problem for a class of uncertain nonlinear systems under
user-defined state constraints.
\subsubsection*{Contributions}\label{subsec:contribution}
The contributions of the present paper are three-fold. First, we formulate the
safety problem as the minimization of the Hamiltonian subject to a constraint
involving the time derivative of the BLF. We subsequently show that the proposed optimization problem is convex, and thus we compute the analytical solution for
the optimal control policy by minimizing the Lagrangian.
Second, we approximate the optimal control policy obtained from the proposed Lagrangian
method and show that this approximate control law renders the system safe for
each time step of the controller run without the help of a backup
stabilizing controller. Third, we extend the ACI
approach\cite{bhasin2013Automatica} to learn the optimal safe control policy online for a general class of uncertain nonlinear systems. Subsequently, we perform
simulation studies on a class of Euler-Lagrange nonlinear systems to show the
efficacy of our proposed methodology. We additionally compare our results with
ACI approach to demonstrate the safety guarantees of the proposed
method.
\subsubsection*{Notations}%
\label{subsec:notation}
 Let $\vect(.)$ denote the vectorization operator of a matrix yielding a column
 vector obtained by stacking the columns of the matrix on top of one another. We
 will use $\nabla$ to denote gradient operator with respect to (w.r.t.) $x$. We use $\norm{.}$ to
 denote the Euclidean norm for vectors and the corresponding induced norm for
 matrices. Let $\mathcal{L}_{\infty}$ denote the set of all bounded signals.
 $\lambda_{\min}(A)$ denotes the minimum eigenvalue of matrix $A$.

\section{Preliminaries}\label{sec:maths}

We consider the following control-affine nonlinear system
\begin{equation}
  \label{eq:systemDynamics}
  \dot{x} = f(x) + g(x)u
\end{equation}
where $x(t) \in \Re^{n}$ is the state vector and $u(t) \in \Re^{m}$ is the
control action. We assume that the state $x(t)$ is measurable. The functions
$f(x) : \Re^{n} \rightarrow \Re^{n}$ and $g(x) : \Re^{n} \rightarrow \R{n}{m}$
are the drift dynamics and control matrix, respectively.

We define the notion of safety as the forward invariance of a
compact set $\C \subset \Re^{n}$ w.r.t the state $x$. In other
words, we deem the system to be safe if
$x(0) \in \C \implies x(t) \in \C \; \fall  t \in \Re_{\ge 0}$. We assume that
the origin is an element of the set $\C$. Additionally, we define the sets
$\partial \C$ and $\text{Int}(\C)$ to be the boundary and the interior of the set
$\C$, respectively.

\begin{assm}
  $f(x)$ and
  $g(x)$ are locally Lipschitz, second-order
  differentiable functions with $f(0)=0$.\label{assm:lipschitz}
\end{assm}

\begin{assm}
  The matrix $g(x)$ has full rank $\fall x \in \C$\label{assm:fullrank}
\end{assm}


\begin{assm}
  The matrix $g(x)$ is known and bounded as
  $\underbar{$g$} < \|g(x)\|< \overline{g}$, where
  $\underbar{$g$}, \overline{g} \in \Re_{>0}$.\label{assm:gBounded}
\end{assm}

We formulate the safe RL problem as the minimization of a cost
functional w.r.t. the control policy $u(t)$, subject to the hard constraint on
the state $x(t)$
\begin{prob}[Constrained Optimal Control]
  \label{prob:constrained}
\begin{subequations}
\begin{align}
 \min_{u(s) \;\fall s \in \Re_{\ge 0}} \hspace{8pt} & \int_{0}^{\infty} r(x(s),u(s)) ds \label{eq:objective}\\
  \text{s.t.} \hspace{20pt} & \dot{x} = f(x) + g(x)u \;\;\;\;\;\;\;\;\;\;\;\; \label{eq:systemConstraint} \\
 & x(t)  \in \C \;\;\fall t \in \Re_{\ge 0} \label{eq:constraint}
\end{align}
\end{subequations}
\end{prob}
where $r:\Re^{n} \times \Re^{m} \rightarrow \Re_{\ge 0}$ is the
instantaneous cost function given by
\begin{equation}
  r(x,u) = Q(x) + \quad{R}{u}
  \label{eq:instantCost}
\end{equation}
where $Q: \C \rightarrow \Re_{\ge 0}$ is
positive-definite cost function in $x$, and
$R \in \Re^{m \times m}$ is positive-definite.

\subsection{Approximate Dynamic Programming}%
\label{subsec:dp}
In the theory of Dynamic Programming, the optimal value function is defined
as
\begin{align}
  V^{*}(x(t)) = \min_{\substack{u(\tau) \\ t \le \tau < \infty}} \int_{t}^{\infty} r(x(s),u(s)) ds
\end{align}
The Hamiltonian of the system is defined as follows
\begin{equation}
  H(x,u,\nabla V) \triangleq r(x,u) + \nabla V^{T} (f(x) + g(x) u)
  \label{eq:perfectHamiltonian}
\end{equation}
We obtain the optimal control law $u^{*}(x)$ for the unconstrained optimal control problem
in \eqref{eq:objective} by minimizing the Hamiltonian w.r.t. the control action $u$
\begin{equation}
  u^{*}(x) = \argmin_{u} H(x,u,\nabla V^{*}) = - \frac{1}{2} R^{-1} g^{T} \nabla V^{*}
  \label{eq:optimalControlLaw}
\end{equation}
Under the optimal control law in \eqref{eq:optimalControlLaw}, the value of
Hamiltonian is identically equal to zero leading to the
Hamilton-Jacobi-Bellman (HJB) equation
\begin{equation}
  H(x,u^{*},\nabla V^{*}) = 0
  \label{eq:perfecthjb}
\end{equation}
The Hamiltonian in \eqref{eq:perfectHamiltonian} can be approximated by replacing $u^{*}$, $V^{*}$, $f(x)$ with
their corresponding estimates $\hat{u}$ (\textbf{Actor}), $\hat{V}$ (\textbf{Critic}) and
$\hat{f}(x)$ (\textbf{Identifier}).
\begin{equation}
  \hat{H}(x,\hat{u},\nabla \hat{V}) \triangleq r(x,\hat{u}) + \nabla \hat{V}^{T} (\hat{f}(x) + g(x) \hat{u})
  \label{eq:estimatedHJB}
\end{equation}
The Bellman Residual error is defined as
\begin{equation}
\begin{aligned}
  \delhjb \triangleq \hat{H}(x,\hat{u},\nabla \hat{V}) -  H(x,u^{*},\nabla V^{*})
\end{aligned}
\end{equation}
We parameterize the Value function via a single layer
neural network (NN)
\begin{equation}
  V^{*}(x) = W^{T} \phi(x) + \ev(x)
  \label{eq:perfectValueFunction}
\end{equation}
where $\phi: \Re^{n}\rightarrow \Re^{p}$ denotes the basis function chosen to
approximate the value function, satisfying $\phi(0) = 0$. The parameter
$W \in \Re^{p}$ denotes the true NN weight and
$\ev: \Re^{n}\rightarrow \Re$ denotes the function approximation error.
\begin{assm}
  The value function approximation error $\ev$ and its derivative w.r.t. state
  are bounded as
  $\norm{\epsilon_{v}(x)} \le \overline{\epsilon},\norm{\nabla \epsilon_{v}(x)} \le \overline{\epsilon}_{d}$.
  Additionally, these bounds approach 0 as the number of neurons approaches infinity.
  \label{assm:actorCriticNNerrorbounded}
\end{assm}
Since the NN weight $W$ is unknown in \eqref{eq:perfectValueFunction}, we maintain two
estimates $\WaHat \in \Re^{p}$ and $\WcHat \in \Re^{p}$ for the control law and the value
function estimate, respectively.

\subsection{BLF-based Constrained Optimal Control Problem}%
\label{subsec:mod}
A positive-definite differentiable function $B_{f}: \C \rightarrow \Re$
satisfying the following properties is called a Barrier Lyapunov function (BLF)
if its time derivative along the system trajectories is negative
semi-definite, i.e. $\dot{B_{f}}(x) \le 0$
\begin{equation*}
  B_{f}(0) = 0, \; \;
  B_{f}(x) > 0 \;\;\fall   x \in \C /\{0\}, \;\;
  \lim_{x\rightarrow \partial \C} B_{f}(x) = \infty
\end{equation*}
The existence of a BLF over $\C$ implies the forward invariance of $\C$
\cite[Lemma 1]{tee2009Automatica}.

\begin{construction}
    $B_{f}(x)$ is constructed in a way such that
    $\exists \; \gamma \in \Re_{>0}$  satisfying $\gamma \norm{\gradBf} \ge B_{f}\; \fall x \in \C$.\label{construction:gradBfupperBound}
\end{construction}
\begin{examp}
 For $x \in \Re$ and $\C = [-1,1]$ a candidate BLF
 $B_{f}(x) = \log(\frac{1}{1-x^{2}})$ with $\gamma=0.5$ satisfies the condition in
 Construction \ref{construction:gradBfupperBound}.
\end{examp}

\begin{rem}
  The constant $\gamma$ would be used to compute the largest attracting subset
  of $\C$.
\end{rem}
Problem \ref{prob:constrained} can be reformulated in terms of BLF as

\begin{prob}
\label{prob:equivalent}
\begin{subequations}
\begin{align}
  \min_{u(s)\;\fall s \in \Re_{\ge 0}} \hspace{8pt} &  H(x,u,\nabla V^{*}) \\
  \text{s.t.} \hspace{20pt} & \frac{d B_{f}}{dt} \Big|_{\dot{x} = f(x) + g(x)u} \le 0 \;\;\;\; \label{eq:barrierConstraint}\\
   & B_{f}(x(0))  < \infty
  \label{eq:modInitCond}
\end{align}
\end{subequations}
\end{prob}
The constraint in \eqref{eq:barrierConstraint} can be rewritten as

\begin{equation}
  \label{eq:barrierConstraintReformulated}
  \gradBf(x)^{T} [f(x) + g(x) u] \le 0
\end{equation}
We observe that the constraint in \eqref{eq:barrierConstraintReformulated} is
affine in the decision variable $u$. This, combined with the fact that the
Hamiltonian in \eqref{eq:perfectHamiltonian} is convex in $u$, makes Problem
\ref{prob:equivalent} a convex optimization problem.
To find an analytical solution, we define the Lagrangian as

\begin{equation}
  L(x,u,\nabla V^{*},\lambda) = H(x,u,\nabla V^{*}) + \lambda \gradBf^{T} (f(x) + g(x)u)
\end{equation}
where $\lambda \in \Re_{\ge 0}$ is the Lagrange multiplier. The control law can
be obtained by minimizing the Lagrangian

\begin{equation}
    u^{*}_{safe}(x,\lambda) = -\frac{1}{2} R^{-1}g^{T}(x)[\nabla V^{*}(x)+\lambda \gradBf(x) ]
  \label{eq:uStarModified}
\end{equation}
\begin{rem}
  The Lagrange multiplier $\lambda$ provides a way to reformulate a constrained
  optimization problem into a weighted unconstrained optimization problem.
  Typically, the expression for Lagrange multipliers are obtained from the KKT
  conditions\cite{almubarak2021CDC}. For simplification of analysis, we
  approximate the optimal Lagrange multiplier with a user-defined constant
  $\lambda$, resulting in a suboptimal solution.
\end{rem}

The estimated safe control law is given by
\begin{equation}
  \hat{u}(x,\lambda) = -\frac{1}{2} R^{-1}g^{T}(x)[ \gradPhi(x)^{T} \WaHat + \lambda \gradBf(x) ]
  \label{eq:finalControlLaw}
\end{equation}
\begin{thm}
  \label{thm:safety}
  Under the control law in \eqref{eq:finalControlLaw} and provided Assumptions
  \ref{assm:lipschitz}-\ref{assm:actorCriticNNerrorbounded} hold, the set $\C$ is forward invariant for the
  system in \eqref{eq:systemDynamics} if $x(0) \in \C$.
\end{thm}
\begin{proof}
  Consider the candidate Lyapunov function as $B_{f}(x) : \C \rightarrow \Re$.
  The time derivative of $B_{f}(x)$ along the trajectories of
  $\dot{x} = f(x) + g(x)\hat{u}$ is given by
  \begin{equation}
    \dot{B}_{f} = \nabla B_{f}^{T} (f(x) + g(x) \hat{u})
  \end{equation}
  Substituting the control law from \eqref{eq:finalControlLaw}, we have
  \begin{equation}
    \dot{B}_{f} = \nabla B_{f}^{T} f - \frac{1}{2} \gradBf^{T} R_{g}
                  \gradPhi^{T} \WaHat - \frac{\lambda}{2} \gradBf^{T} R_{g} \gradBf
      \label{eq:Bdot}
  \end{equation}
  where we define $R_{g}(x) \triangleq g(x)R^{-1}g^{T}(x)$ and
  $R_{s}(x) \triangleq \gradPhi(x) R_{g}(x) \gradPhi^{T}(x)$. Under Assumption
  \ref{assm:fullrank}, $R_{g}(x)$ is positive-definite. Additionally, $R_{g}$ is
  bounded as $\norm{R_{g}(x)} \le \overline{R}_{g} \;\fall x \in \C$. Since
  $f(x)$ and $\gradPhi(x)$ are continuous functions over compact set $\C$,
  $\norm{f(x)} \le \overline{f}, \norm{\gradPhi(x)} \le \overline{\phi}_{d}\;\fall x \in \C$.
  We can upper bound the right hand side of \eqref{eq:Bdot} by
  \begin{equation}
      \dot{B}_{f} \le (\overline{f} + \frac{1}{2} \overline{\phi}_{d}
     \overline{W}_{a}  \overline{R}_{g} ) \norm{\gradBf} - \frac{\lambda}{2} \lambda_{\min}(R_{g})\norm{\gradBf}^{2}
  \end{equation}
  where $\overline{W}_{a} \in \Re_{>0}$ is the bound on the true NN weight $W$
  which is subsequently enforced on $\WaHat$ via a projection operator\cite{lavretsky2011Arxiv}. We observe that the $\dot{B}_{f}$ is negative outside the compact set $\Omega
    = \{ x \in \Re^{n} :\norm{\gradBf} \le \overline{B}_{d}
       \}$, where $\overline{B}_{d} \triangleq \frac{\overline{f}
 + \frac{1}{2} \overline{\phi}_{d}
  \overline{W}_{a}  \overline{R}_{g}}{\frac{\lambda}{2}\lambda_{\min}(R_{g})}$
is a computable finite positive constant. Under the condition in Construction
\ref{construction:gradBfupperBound} we can upper bound the value of Barrier function as
\begin{equation}
  \label{eq:4}
  B_{f}(x(t)) \le \max\big(B_{f}(x(0)), \gamma \overline{B}_{d} \big)
\end{equation}
Since $x(0) \in \C$, the $B_{f}(x(0))$ is finite. Thus,
$B_{f}(t) \in \Linf$. Since the value of the Barrier function along the system
trajectory is bounded, then by the definition of $B_{f}(x)$, at no point in
time, the state trajectory intersects the boundary of the safe set $\partial \C$
\cite[Lemma 1]{tee2009Automatica}. Thus the state
$x(t) \in \C \;\;\fall t \in \Re_{\ge 0}$ and the system is forward invariant.
Since the BLF is continuously differentiable in $x$, the $\gradBf(x)$ is a
continuous function over the compact set $\Omega$. Thus,
$\norm{\gradBf} \in \Linf$. Since all constituents of the control
law in \eqref{eq:finalControlLaw} are bounded, we can conclude that $\hat{u}(t)\in \Linf$.
\end{proof}
\begin{rem}
 Theorem \ref{thm:safety} proves that the control policy in
 \eqref{eq:finalControlLaw} guarantees safety for all time. Further, the control
 policy doesn't switch between a stabilizing backup policy and the RL policy,
 which is a distinct advantage over approaches that rely on an elusive backup policy.
\end{rem}

\subsection{Actor-Critic Design}%
\label{subsec:actorCriticMaths}
The actor NN weight $\WaHat$ and the critic NN weight $\WcHat$ are updated to
minimize the norm of the estimation errors $\WcTilde \triangleq W - \WcHat$ and
$\WaTilde \triangleq W - \WaHat$.
A least-squares update law for the critic can be obtained from the consideration
of the integral squared Bellman error
\cite{bhasin2013Automatica} as follows
\begin{equation}
  \label{eq:integralBellmanError}
  E_{c} = \int_{0}^{t} \delhjb^{2}(\tau) d\tau
\end{equation}
Defining $\omega \triangleq \frac{\partial \delhjb}{\partial
\WcHat}$ the update law for critic is given as
\begin{equation}
  \label{eq:criticUpdateLaw}
    \dot{\hat{W}}_{c} = \eta_{c} \Gamma \frac{\omega}{1+\nu \omega^{T}\Gamma\omega}\delta_{hjb}
\end{equation}
where the learning rate $\eta_{c}$ and normalizing factor $\nu$
are positive user-defined constants.
The positive-definite covariance matrix $\Gamma \in \R{p}{p}$ is updated via the update law
\begin{equation}
  \label{eq:gammaUpdateLaw}
  \dot{\Gamma} = \beta \Gamma - \eta_{c} \Gamma \frac{\omega \omega^{T}}{1+\nu \omega^{T}\Gamma\omega} \Gamma
\end{equation}
Under the aforementioned update law of the covariance matrix, the
following bounds can be established
\begin{equation}
  \label{eq:gammaBounds}
  \varphi_{1}I_{p} \preccurlyeq \Gamma(t) \preccurlyeq \varphi_{0} I_{p} \;\;\; \fall t \ge 0
\end{equation}
where $\preccurlyeq$ denotes the semi-definite ordering and
$\varphi_{0} > \varphi_{1}$ are positive constants.
The update law for the actor is obtained by the gradient descent of the cost
function in \eqref{eq:integralBellmanError}
\begin{equation}
  \label{eq:actorUpdateLaw}
  \begin{aligned}
    \dot{\hat{W}}_{a} &= \proj\Big[ -\frac{\eta_{a1}}{\sqrt{1+\omega^{T}\omega}} R_{s}(\WaHat - \WcHat)\delta_{hjb} \\
                      &\;\;-\eta_{a2}(\WaHat - \WcHat) - \frac{1}{2}\lambda \gradPhi R_{g}\gradBf \Big]
  \end{aligned}
\end{equation}
where the projection operator, $\proj(.)$ \cite{lavretsky2011Arxiv} is used to keep the
estimates of the actor parameter bounded. The positive constants
$\eta_{a1}, \eta_{a2} \in \Re_{> 0 }$ user defined gains. The last two terms in
the argument of the projection operator are attributed to the subsequent
Lyapunov analysis in Subsection \ref{subsec:lyap}.

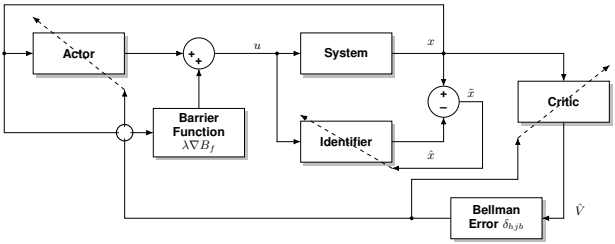
\begin{figure}[h]
\centering
\resizebox{0.95\linewidth}{!}{
\begin{tikzpicture}[font=\small\sffamily\bfseries,node distance =
3cm, scale=1,every node/.style={transform shape}]
\node[draw,circle,minimum size=0.8cm,fill=white] (sum) at (0,0){};
\node [frame,below=1cm of sum] (gradB){Barrier Function $\lambda \nabla B_{f}$};
\node [frame,left=1.5cm of sum] (actor) {Actor};
\node [frame,right=2.2cm of sum] (plant) {System};
\node [frame,below=1.2cm of plant] (ident) {Identifier};
\coordinate[right=2.5cm of plant.south](P);
\coordinate[right=2.5cm of plant.center](x);
\node[left=-1pt] at (sum.center){\textbf{+}};
\node[below] at (sum.center){\textbf{+}};
\node[draw,circle,minimum size=0.8cm,fill=white,below=0.3cm of
P] (error){};
\node[above] at (error.center){\textbf{+}};
\node[below] at (error.center){\textbf{--}};
\node [frame,right=1.5cm of error] (critic) {Critic};
\node [frame,below right=0.9cm and 1.5cm of ident] (bellman)
{Bellman Error $\delta_{hjb}$};
\coordinate[left=0.6cm of plant.west](uBreak);
\coordinate[right=1cm of plant.east](xMark);
\coordinate[right=1cm of ident.east](xHatMark);
\draw[line] (actor) -- (sum);
\draw[line] (gradB) -- (sum);
\draw[line] (sum) -- node[midway,above]{$u$} (plant);
\draw[line] (uBreak) |- (ident);
\draw[line] (plant)  -| (error);
\node[above=1mm of xMark,align=left]{$x$};
\node[below=1mm of xHatMark,align=left]{$\hat{x}$};
\draw[line] (ident) -| (error);
\draw[line] (x) -- ++ (0,1.25) -- (-5,1.25) |- (actor);
\draw[line,name path=firstPath] (-5,0) |- (gradB);
\coordinate[below right=0.15cm and 0cm of ident] (i1);
\coordinate[above left=0.15cm and 0cm of ident] (i2);
\coordinate[below right=0.4cm and 0cm of actor] (a1);
\coordinate[above left=0.4cm and 0cm of actor] (a2);
\coordinate[below left=0.4cm and 0cm of critic] (c1);
\coordinate[above right=0.4cm and 0cm of critic] (c2);
\draw[line] (error) -- ++ (1,0) node[midway,above]{$\tilde{x}$}  |-
(i1);
\draw[line,dashed] (i1) -- (i2);
\draw[line] (x) -| (critic);
\coordinate[left=1cm of bellman] (b);
\node[circle,fill,inner sep=1pt](m) at (x){};
\node[circle,fill,inner sep=1pt](m) at (-5,0){};
\node[circle,fill,inner sep=1pt](m) at (uBreak){};
\node[circle,fill,inner sep=1pt](m) at (b){};
\draw[line] (critic) node[below left=2.5cm and 0.5cm of critic.east]{$\hat{V}$} |- (bellman);
\draw[line,name path=secondPath] (bellman) -| (a1);
\path [name intersections={of= firstPath and
secondPath,by=inter}];
\node[circle,fill=white,line width=1mm,inner sep=3pt](m) at (inter){};
\node[circle, draw=black, thick, inner sep=0pt, minimum size=12pt] (1) at (inter) {};
\draw[line,dashed] (a1) -- (a2);
\draw[line,dashed] (c1) -- (c2);
\draw[line] (b) -- ++(0,1) -| (c1);
\end{tikzpicture}
}
\caption{Block diagram of the proposed Reinforcement Learning algorithm}%
\label{fig:blockDiag}
\end{figure}

\subsection{Identifier Design}%
\label{subsec:identifier}
We represent the system drift dynamics $f(x)$ in \eqref{eq:systemDynamics}  via a
two-layer NN parameterized by $\Wf \in \R{l}{n}$ and
$\Vf \in \R{n}{l}$. We represent the activation function of the NN by
$\sigma: \Re^{l}\rightarrow \Re^{l} $. The dynamics of the system can be written as
\begin{align}
  \dot{x} = \WfT \sig + \ef + g(x) \tau
\end{align}
The following state estimator is designed by involving estimates of $W_{f}$ and $V_{f}$ in
the form of $\WfHat$ and $\VfHat$ respectively
\begin{equation}
  \dot{\hat{x}} = \WfHatT \sigHat + g(x) \tau + k \tilde{x}
  \label{eq:identifierderivative}
\end{equation}
where $\tilde{x} \triangleq x- \hat{x}$ denotes the state estimation error and
$k \in \Re_{>0}$ is a feedback gain.
\begin{assm}
  The parameters $W_{f}, V_{f}$ are assumed to be bounded and
  $\|\sigma(\cdot)\|<\overline{\sigma},\|\nabla \sigma(\cdot)\| < \overline{\sigma}_{d} \; \; \fall x \in \C$.\label{assm:identifierbound}
\end{assm}

Based on a Lyapunov analysis (omitted here in the interest of space), we
design the following adaptive laws for the NN parameters
\begin{equation}
  \label{eq:identifierAdaptiveLaws}
  \dot{\hat{W}}_{f}  = \proj(\Gamma_{wf}\hat{\sigma}\tilde{x}^{T}) \;,\;
  \dot{\hat{V}}_{f}  = \proj(\Gamma_{vf}x \tilde{x} ^{T} \WfHatT \nabla \hat{\sigma})
\end{equation}
where $\Gamma_{wf} \in \R{l}{l}$  and $\Gamma_{vf} \in \R{n}{n}$ are
positive definite gain matrices. We define $\WfTilde \triangleq W_{f} - \WfHat$
and $\VfTilde \triangleq V_{f} - \VfHat$.
\begin{thm}
  \label{thm:ident}
  Under the identifier update laws given by \eqref{eq:identifierderivative}, \eqref{eq:identifierAdaptiveLaws} and Assumption \ref{assm:identifierbound},
  the state identification error ($\tilde{x}(t)$), the error in NN
  parameters ($\tilde{W}_{f}(t)$ and $\tilde{V}_{f}(t)$) are Uniformly Ultimately
  Bounded (UUB)

\end{thm}
\begin{proof}
(Sketch) We define an auxiliary state
$\zeta \triangleq { [\tilde{x}^{T},\vect(\WfTilde)^{T}, \vect(\VfTilde)^{T}] }^{T}$.
Considering the following Lyapunov function
  \begin{equation*}
    V_{1}(\zeta) = \frac{1}{2}\tilde{x}^{T}\tilde{x} +
    \frac{1}{2}tr(\Wfquad) +
    \frac{1}{2}tr(\Vfquad)
  \end{equation*}
  One can show that $\dot{V}_{1}(\cdot)$ is negative whenever $\zeta$ lies outside
  the compact set $\Omega_{\zeta} \triangleq \{\zeta : \norm{\tilde{x}} \le  \frac{\overline{\sigma}^{2}  \norm{\WfTilde}^{2}}{4k^{2}}  + \frac{1}{k}\overline{\chi}
\}$, where $\overline{\chi}$ is the computable upper bound of $\ef$ and higher
order terms originating from Taylor's approximation of $\sigma$. Hence the state $\zeta$ is UUB.
\end{proof}
Block diagram of the resulting system is shown in Fig. \ref{fig:blockDiag}

\subsection{Stability analysis}%
\label{subsec:lyap}
The Bellman estimation error can be written in its unmeasurable form as
\begin{equation}
  \delhjb = \nabla \hat{V}^{T} \hat{F}_{\hat{u}} + r(x,\hat{u}) - \nabla V^{*T}
  {F}_{u^{*}} - r(x,{u^{*}})
  \label{eq:unmeasurableDeltaHJB}
\end{equation}
where $F_{u^{*}} \triangleq f(x) + g(x) u^{*}$ and
$\hat{F}_{\hat{u}} \triangleq \hat{f}(x) + g(x) \hat{u}$. Additionally, we
define $\tilde{F}_{\hat{u}} \triangleq F_{u^{*}} - \hat{F}_{\hat{u}}$.

Substituting the instantaneous cost from \eqref{eq:instantCost}, and the
NN approximations of $V^{*}$  from \eqref{eq:perfectValueFunction}
and its estimate $\hat{V}$  we have
\begin{equation}
  \delhjb = \WcHat^{T} \omega - [W^{T}\gradPhi + \nabla \ev^{T}] F_{u^{*}} + \quad{R}{\hat{u}} -
  u^{*T} R u^{*}
  \label{eq:unmeasurableDeltaHJBPart2}
\end{equation}
Substituting optimal control $u^{*}$ from \eqref{eq:uStarModified} and its
estimate $\hat{u}$ from \eqref{eq:finalControlLaw} in
  \eqref{eq:unmeasurableDeltaHJBPart2} and simplifying, we have
\begin{equation}
  \delhjb = -\WcTilde^{T} \omega + T_{1}
  \label{eq:unmeasurableDeltaHJBFinal}
\end{equation}
where
\begin{equation*}
  \begin{aligned}
    T_{1} \triangleq &- W^{T} \gradPhi \tilde{F}_{\hat{u}}
      - \nabla \ev^{T} F_{u^{*}}
      + \frac{1}{4} \quad{R_{s}}{\WaHat}
      - \frac{1}{4}\quad{R_{s}}{W} \\
    &-\frac{1}{4} \quad{R_{g}}{\nabla \ev}
      - \frac{1}{2}\lambda \gradBf^{T} R_{g}(\gradPhiT \WaTilde + \nabla \ev) \\
    &- \frac{1}{2} W^{T} \gradPhi R_{g} \nabla \ev
  \end{aligned}
\end{equation*}
Substituting \eqref{eq:unmeasurableDeltaHJBFinal} into the dynamics of the
critic estimation error $\WcTildeDot = - \dot{\hat{W}}_{c}$ we obtain two components, a nominal dynamics term ($\Omega_{nom}$) and a perturbation
term ($\Delta$)
\begin{equation}
  \WcTildeDot = \underbrace{-\eta_{c} \Gamma \psi \psi^{T} \WcTilde}_{\Omega_{nom}} + \underbrace{\eta_{c} \Gamma \frac{\omega}{1+ \nu \quad{\Gamma}{\omega}} T_{1}}_{\Delta}
  \label{eq:nomPlusPert}
\end{equation}
where
$\psi(t) \triangleq \frac{\omega(t)}{\sqrt{1+ \nu \quad{\Gamma(t)}{\omega(t)}}} \in \Re^{n}$
is the normalized gradient vector for the update law of the critic. The
regressor $\psi(t)$ is bounded as
\begin{equation}
  \label{eq:3}
  \norm{\psi(t)} \le \frac{1}{\sqrt{\nu \varphi_{1}}} \;\; \fall t \ge 0
\end{equation}
The nominal dynamics
$\dot{\tilde{W}}_{c} = \Omega_{nom}$ is globally exponentially stable
 (GES), provided that the bounded signal $\psi(t)$ is persistently exciting (PE)\cite{bhasin2013Automatica}.
Consequently, there exists a
positive-definite scalar-valued function $V_{c} (\WcTilde, t)$ such that the
following conditions are satisfied
\begin{equation}
  \label{eq:definitionOfVc}
  \begin{aligned}
   c_{1} \norm{\WcTilde}^{2} \le V_{c}(\WcTilde,t) &\le c_{2} \norm{\WcTilde}^{2} \\
    \frac{\partial V_{c}}{\partial t} + \frac{\partial V_{c}}{\partial \WcTilde} \Omega_{nom} &\le -c_{3}\norm{\WcTilde}^{2} \\
 \norm{\frac{\partial V_{c}}{\partial \WcTilde}} &\le c_{4} \norm{\WcTilde}
  \end{aligned}
\end{equation}
where $c_{1},c_{2},c_{3},c_{4}$ are positive scalar constants. Additionally, we
define the following term that would appear in the subsequent Lyapunov analysis

\begin{equation*}
  \begin{aligned}
    T_{2} &\triangleq \frac{1}{4} \quad{R_{g}}{\nabla \ev}
    - \frac{1}{2} \lambda \gradBf^{T} R_{g} \nabla \ev
    + \frac{1}{2}\WaTilde^{T} \gradPhi R_{g} \nabla \ev \\
    &- \frac{1}{4} \quad{R_{s}}{\WaHat}
    + \lambda \gradBf^{T} f(x)
    - \lambda \gradBf^{T} R_{g} \gradPhi^{T} \WaHat
  \end{aligned}
\end{equation*}
Under the Assumptions \ref{assm:gBounded}-\ref{assm:actorCriticNNerrorbounded}, Theorems
\ref{thm:safety}-\ref{thm:ident}, we can obtain the following computable
bounds
\begin{equation}
  \label{eq:Bounds}
  \begin{aligned}
    \norm{\WaTilde} &\le \WaBar,
    \norm{T_{1}} &\le \ToneBar,
    \norm{R_{s}} &\le \RsBar,
    \norm{T_{2}} &\le \TtwoBar
  \end{aligned}
\end{equation}
where $\WaBar, \ToneBar, \RsBar, \TtwoBar \in \Re_{>0}$ are computable positive constants. Subsequently, we can bound the perturbation term in \eqref{eq:nomPlusPert} by
\begin{equation}
  \label{eq:perturbationBound}
  \norm{\Delta} \le \frac{\eta_{c}\varphi_{0} \ToneBar}{2\sqrt{\nu \varphi_{1}}}
\end{equation}
where $\varphi_{1}$ was defined in \eqref{eq:gammaBounds}.
\begin{thm}
 Provided Assumptions \ref{assm:lipschitz}-\ref{assm:identifierbound} hold, the regressor matrix $\psi(t)$
 is PE, and the following gain conditions
 $c_{3} < \eta_{a1}\WaBar \RsBar $ and  $4\eta_{a2} > \RsBar$ are satisfied,
 then the control action in \eqref{eq:finalControlLaw}, the actor and critic
 update laws from \eqref{eq:criticUpdateLaw} and \eqref{eq:actorUpdateLaw}, and the
 identifier \eqref{eq:identifierderivative}, \eqref{eq:identifierAdaptiveLaws}
 guarantee that the state $x(t)$, actor weight estimation error $\WaTilde(t)$
 and the critic weight estimation error $\WcTilde(t)$ are UUB. 
\end{thm}

\begin{proof}
  We define the auxiliary state $z \triangleq
  [x^{T} , \vect(\WcTilde)^{T} , \vect(\WaTilde)^{T}]^{T}$.  We consider the
  following locally positive-definite candidate Lyapunov
  function
  \begin{align}
    V_{L}(z,t) &= V^{*}(x)+ \lambda B_{f}(x) + V_{c}(\WcTilde,t) +  \frac{1}{2} \quad{ }{\WaTilde}
  \end{align}
  Computing the derivative of the same w.r.t. time along the system trajectory
  \begin{equation}
  \begin{aligned}
   \dot{V}_{L}(z,t) &= ( \nabla V^{*} + \lambda \nabla B_{f})^{T} [f + g\hat{u}] + \frac{dV_{c}}{dt}   - \WaTilde^{T} \dot{\hat{W}}_{a}
  \end{aligned}
  \end{equation}
  Using \eqref{eq:perfecthjb}, \eqref{eq:definitionOfVc} we have
  \begin{equation}
  \begin{aligned}
    \dot{V}_{L}  \le & -Q(x) - u^{*T}R u^{*} - \nabla V^{*T} g \tilde{u} + \lambda \nabla B_{f}^{T} [f + g\hat{u}] \\
    & - c_{3} \norm{\WcTilde} ^{2} + c_{4} \norm{\WcTilde}\norm{\Delta} - \WaTilde^{T} \dot{\hat{W}}_{a}
  \end{aligned}
  \end{equation}
  Substituting the control law from \eqref{eq:finalControlLaw}, bounds from
 \eqref{eq:Bounds} - \eqref{eq:perturbationBound}, the actor update law from
 \eqref{eq:actorUpdateLaw}, the $\delhjb$ from
 \eqref{eq:unmeasurableDeltaHJBFinal}  and using the properties of the $\proj(.)$ operator
 \cite{lavretsky2011Arxiv} we have
 \begin{equation}
   \begin{aligned}
        \dot{V}_{L} \le  & -Q(x)
        -(c_{3} - \eta_{a1} \WaBar \RsBar)\norm{\WcTilde}^{2}
        - (\eta_{a2} - \frac{\RsBar}{4}) \norm{\WaTilde}^{2} \\
        &- \frac{3 \lambda^{2}}{4} \lambda_{\min}(R_{g})\norm{\gradBf}^{2}
          + \TtwoBar + \eta_{a1} \WaBar^{2} \RsBar \ToneBar + T_{3} \norm{\WcTilde}
   \end{aligned}
 \end{equation}
 where $T_{3} \triangleq \frac{c_{4} \eta_{c} \varphi_{0}\ToneBar}{2 \sqrt{\nu \varphi_{1}}}
          + \eta_{a2} \WaBar + \eta_{a1} \WaBar \RsBar \ToneBar + \eta_{a1} \WaBar^{2} \RsBar$. Under the gain condition of $c_{3} >  \eta_{a1} \WaBar \RsBar$, completing the
 squares yields
 \begin{equation}
   \begin{aligned}
        \dot{V}_{L} \le & -Q(x)
        -(1-\theta)(c_{3} - \eta_{a1} \WaBar \RsBar)\norm{\WcTilde}^{2} \\
        &- (\eta_{a2} - \frac{\RsBar}{4}) \norm{\WaTilde}^{2}
        - \frac{3 \lambda^{2}}{4} \lambda_{\min}(R_{g})\norm{\gradBf}^{2} \\
        &+ \frac{T_{3}^{2}}{4\theta (c_{3} - \eta_{a1} \WaBar \RsBar )}
        + \TtwoBar + \eta_{a1} \WaBar^{2} \RsBar \ToneBar
   \end{aligned}
 \end{equation}
 where $\theta \in (0,1)$. Under the additional gain condition of
 $4\eta_{a2} > \RsBar$ , there exist two class $\mathcal{K}$ functions $\alpha_{1}$
 and $\alpha_{2}$ such that the following inequalities hold
 \begin{equation}
   \begin{aligned}
     &\alpha_{1}(\norm{z}) \le Q(x) + (1-\theta) (c_{3} - \eta_{a1} \WaBar \RsBar) \norm{\WcTilde}^{2} \\
     &+ (\eta_{a2} - \frac{\RsBar}{4}) \norm{\WaTilde}^{2}
    + \frac{3 \lambda^{2}}{4} \lambda_{\min}(R_{g})\norm{\gradBf}^{2} \le \alpha_{2}(\norm{z})
   \end{aligned}
 \end{equation}
 The derivative of Lyapunov function is upper-bounded by
 \begin{equation}
   \begin{aligned}
        \dot{V}_{L} \le & -\alpha_{1}(\norm{z}) + \frac{T_{3}^{2}}{4\theta (c_{3} - \eta_{a1} \WaBar \RsBar )}
          + \TtwoBar + \eta_{a1} \WaBar^{2} \RsBar \ToneBar
   \end{aligned}
 \end{equation}
 we observe that $\dot{V}_{L}(z,t)$ is negative whenever $z(t)$ lies outside the
 compact set $\Omega_{z} \triangleq \{z : \norm{z} \le \alpha_{1}^{-1}(\frac{T_{3}^{2}}{4\theta (c_{3} - \eta_{a1} \WaBar \RsBar )}
          + \TtwoBar + \eta_{a1} \WaBar^{2} \RsBar \ToneBar)\}$. We can thus
          conclude that the norm of the auxiliary state $\norm{z(t)}$ is
          UUB.
\end{proof}

\section{Simulation Results}%
\label{sec:sim}

To test the efficacy of the proposed control law, we perform a simulation study
on a class of nonlinear Euler-Lagrangian systems
\begin{equation}
  \label{eq:ELsys}
  \begin{aligned}
  M(q)\ddot{q} + C_{m}(q,\dot{q}) \dot{q} + G(q)+F_{d}(\dot{q}) = \tau(t)
  \end{aligned}
\end{equation}

Specifically, we consider the safe, optimal control problem for a two-link robot
manipulator system
\begin{equation*}
  \begin{aligned}
    M(q)  &= \begin{bmatrix}
         p_1 + 2 p_3 c_2 & p_2 + p_3 c_2 \\
         p_2 + p_3 c_2 & p_2 \\
             \end{bmatrix} \; , \;
    F_{m}(\dot{q})  = \begin{bmatrix}
        f_{d_1} \dot{q}_1 \\ f_{d_2} \dot{q}_2
    \end{bmatrix} \\
    C_{m}(q,\dot{q})  &= \begin{bmatrix}
         -p_3 s_2 \dot{q}_2 & -p_3 s_2 (\dot{q}_1 + \dot{q}_2) \\
         p_3 s_2 \dot{q}_1 & 0
       \end{bmatrix} \; ,\; G(q) = 0_{2 \times 1}
  \end{aligned}
\end{equation*}
where the signals $q_{1}(t),q_{2}(t) \in \Re$ denote the angular position of
the two link joints in radians. The parameters used for the simulation are
       $p_1 = \SI{3.473}{kg.m}  $,
       $p_2 = \SI{0.196}{kg.m}  $,
       $p_3 = \SI{0.242}{kg.m}  $,
       $f_{d_{1}} = \SI{5.3}{N.s}  $,
       $f_{d_{2}} = \SI{1.1}{N.s}  $.

The system is then reformulated to the control affine form given in
\eqref{eq:systemDynamics} by defining the system state as
$x = [q_{1},q_{2},\dot{q}_{1},\dot{q}_{2}]^{T}$ and the control action as $u = \tau$. We seek to solve the optimal
control problem (Problem \ref{prob:constrained}) considering the following the
cost function components as
$Q(x) = \quad{}{x}$ and $ R = \mathbb{I}_{2\times2} $ and the state constraint set
$\C = \{x \in \Re^{4}: |x_{i}|< a_{i}\;\fall i \in \{1,2,3,4\} \}$
\footnote{We consider rectangular constraints for the ease of visualization.
  The proposed method can be easily extended to consider other types of state constraints.}.
We consider the following candidate Barrier
Lyapunov Function:  $B_{f} = \sum_{i = 1}^{n} \log{\frac{a_{i}^{2}}{a_{i}^{2} - x_{i}^{2}}}$.
For the given two-link robot manipulator system, we have considered
$a_{i} = 5 \;\fall i \in \{1,2,3,4\}$.
We observe that there exists a $\gamma = 5$ that satisfies the condition
$\gamma \norm{\gradBf} > B_{f}$.

The Critic NN and the Identifier NN were considered to be two-layer
NNs with sigmoidal activation function and hidden layer consisting of 30 and
5 neurons respectively.
The gains for the actor-critic components were chosen as $\eta_{c} = 2$ ,
$\eta_{a1} = 1$ and $\eta_{a2} = 50$. The forgetting factor $\beta = 0.001$ and
the multiplier $\nu = 5$. The Lagrangian multiplier $\lambda$ was set to 100 to
ensure that the value of the bound $\overline{B}_{d}$ is of a reasonable magnitude. For identifier, we chose the gains
$\Gamma_{wf}= 10 \mathbb{I}_{l\times l}$ ,
$\Gamma_{vf}= 10 \mathbb{I}_{n \times n}$. The
identifier feedback gain was set to $k = 10$.
The covariance matrix was initialized to $\Gamma(0) = \mathbb{I}_{p \times p}$
and all the NN weights were initialized in the range of $[-1,1]$
with a uniform probability distribution.

\begin{figure}[htpb]
  \centering
  \subfloat[State Trajectory\label{fig:state}]{\ifig{0.5}{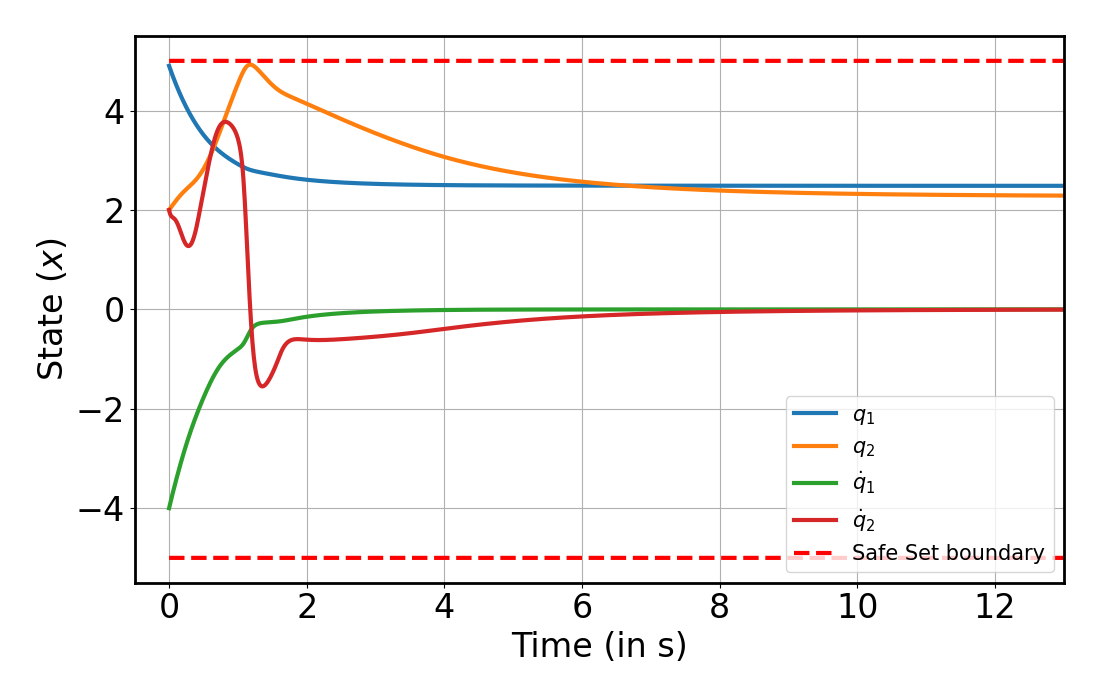}}
  \hfill
  \subfloat[Control Effort\label{fig:u}]{\ifig{0.5}{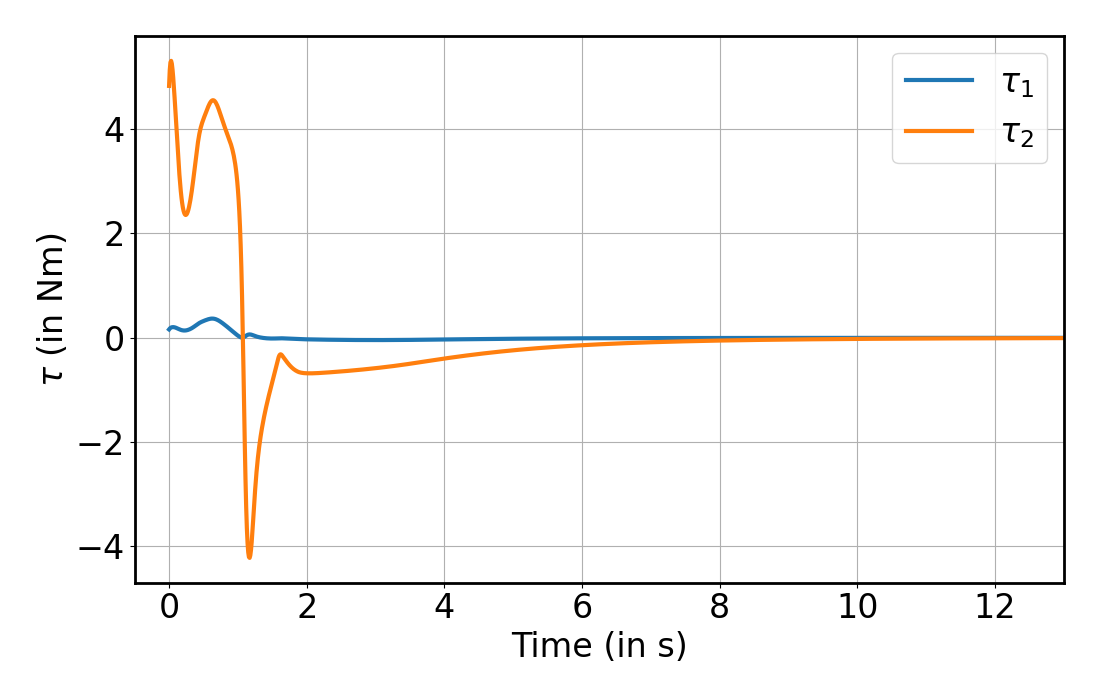}} \\
  \subfloat[Identifier Estimation Error\label{fig:error}]{\ifig{0.5}{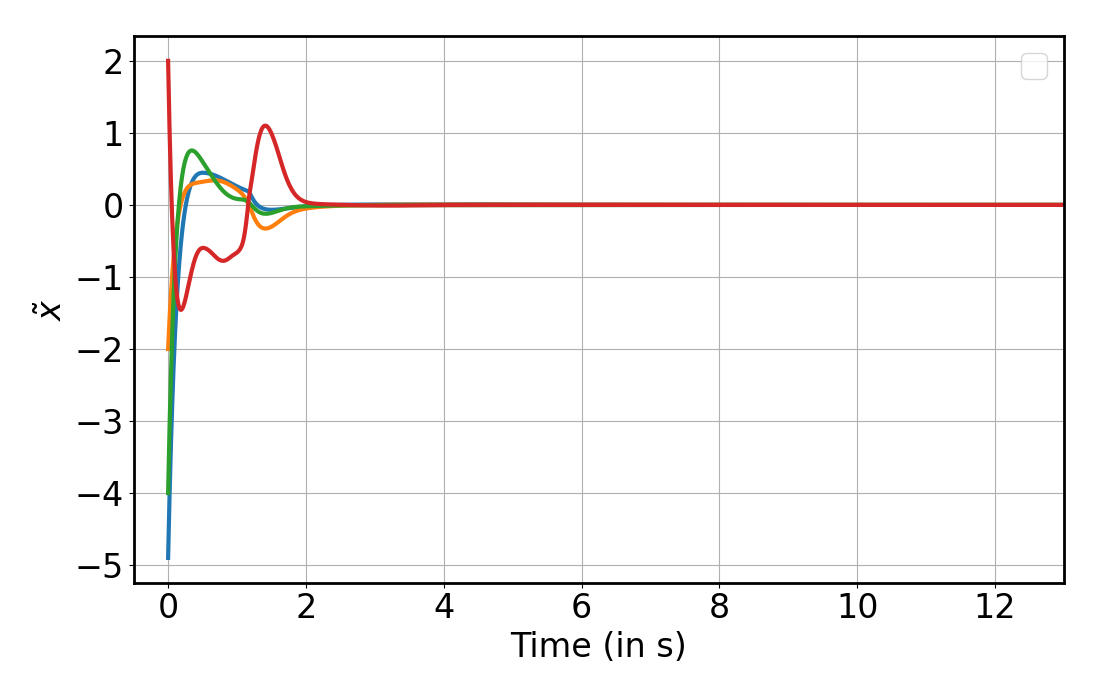}}
  \hfill
  \subfloat[Comparision with Unconstrained ACI\cite{bhasin2013Automatica}\label{fig:comp}]{\ifig{0.5}{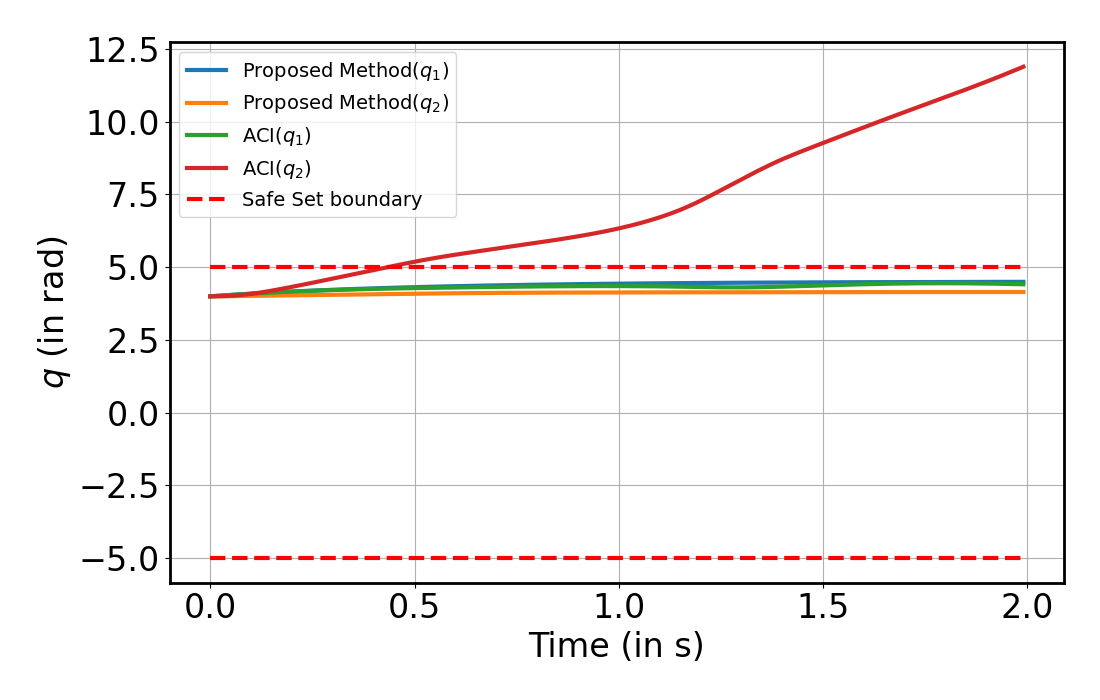}}
  \caption{Simulation Plots of our proposed method}
\end{figure}

Fig. \ref{fig:state} shows the state trajectory of the system under the influence
of the proposed control law. We observe that all of the states
are inside the prescribed limit shown in red dotted lines. Additionally, the state remains uniformly
ultimately bounded. Fig. \ref{fig:u} shows the control effort imposed by the
controller. Fig. \ref{fig:error} shows the estimation error of the identifier.
It can be seen that the estimation error converges very close to zero. In the
Fig. \ref{fig:comp} we observe a comparison of performance in the initial 2
seconds of training of the proposed method with the ACI method
\cite{bhasin2013Automatica}. The hyper-parameters for the algorithm outlined in
\cite{bhasin2013Automatica}, were taken in similar orders of magnitude as
detailed in that article to enable a juxtaposition of the two results for better comparison.
We observe that while the ACI method initially violates the safety criterion,
the proposed method manages to keep the states well within the boundaries of the
safe set, highlighting the transient safety guarantees of the proposed method.

\end{document}


\section{Conclusions and Future Work}
We develop an online Actor-Critic-Identifier
architecture-based safe RL algorithm to solve the optimal regulation problem for
a class of uncertain
nonlinear systems while adhering to
user-defined state constraints. We formulate the safety problem as a convex
optimization problem involving the minimization of the Hamiltonian subject to
the negative semi-definiteness of a candidate BLF. We derive an optimal control
law for the constrained system by solving the Lagrangian and show that the
on-policy RL algorithm ensures the forward invariance of the constraint set
without the need to switch to an external stabilizing backup controller. We
subsequently develop adaptation laws to learn the optimal policy and
demonstrate that all closed-loop signals are UUB. Finally, we demonstrate the
effectiveness of our controller on a two-link robot manipulator system and
compare our results with that of the existing literature. We show that the
proposed method successfully managed to ensure safety during the initial phase
of training while the existing approach shows safety violations.
Future work includes extending the proposed methodology to include actuation
constraints as well as state constraints, without
requiring a backup controller. The optimality of the proposed
controller may be improved by considering Lagrange multiplier obtained from KKT conditions.

\addtolength{\textheight}{-12cm}





\bibliographystyle{jabbrv_IEEEtran}
\bibliography{output.bib}

\end{document}